\documentclass[twocolumn,showpacs,amsmath,amssymb,prd,epsf]{revtex4}

\usepackage{graphicx}
\usepackage{graphicx}
\usepackage{dcolumn}
\usepackage{bm}
\usepackage{epsf}

\def\agt{\mathrel{\raise.3ex\hbox{$>$}\mkern-14mu\lower0.6ex\hbox{$\sim$}}}
\def\alt{\mathrel{\raise.3ex\hbox{$<$}\mkern-14mu\lower0.6ex\hbox{$\sim$}}}

\newcommand{\beq}{\begin{equation}}
\newcommand{\eeq}{\end{equation}}
\newcommand{\beqn}{\begin{eqnarray}}
\newcommand{\eeqn}{\end{eqnarray}}

\begin{document}

\title{Gravitational waves and neutrino emission from 
the merger of binary neutron stars}

\author{Yuichiro Sekiguchi}

\author{Kenta Kiuchi}

\author{Koutarou Kyutoku}

\author{Masaru Shibata}

\affiliation{Yukawa Institute for Theoretical Physics, 
Kyoto University, Kyoto 606-8502, Japan
}

\begin{abstract}
Numerical simulations for the merger of binary neutron stars are
performed in full general relativity incorporating a
finite-temperature (Shen's) equation of state (EOS) and neutrino cooling for
the first time. It is found that for this stiff EOS, a
hypermassive neutron star (HMNS) with a long lifetime ($\gg 10$ ms) is
the outcome for the total mass $\alt 3.0M_{\odot}$. It is shown that
the typical total neutrino luminosity of the HMNS is $\sim 3$--$
8\times 10^{53}$~erg/s and the effective amplitude of gravitational
waves from the HMNS is 4--$6 \times 10^{-22}$ at $f=2.1$--2.5~kHz for
a source distance of 100~Mpc. We also present the neutrino luminosity
curve when a black hole is formed for the first time.
\end{abstract}

\maketitle

{\em Introduction}: Coalescence of binary neutron stars (BNS) is one
of the most promising sources for next-generation kilo-meter-size
gravitational-wave (GW) detectors~\cite{LIGOVIRGO}, and also 
a possible candidate for the progenitor of short-hard gamma-ray bursts
(SGRB)~\cite{GRB-BNS}. Motivated by these facts, numerical simulations
have been extensively performed for the merger of BNS in the framework
of full general relativity in the past decade since the first success
in 2000~\cite{SU00} (see, e.g., \cite{Duez} for a review).

BNS evolve due to gravitational radiation reaction and eventually
merge. Before the merger sets in, each neutron star is cold (i.e.,
thermal energy of constituent nucleons is much smaller than the Fermi
energy), because 
thermal energy inside the neutron stars is significantly reduced by 
neutrino and photon coolings due to the long-term evolution (typically $\agt 10^8$~yrs) 
until the merger~\cite{ST83}. 
By contrast, after the merger, shocks are generated by hydrodynamic
interactions. In particular, when a hypermassive neutron star (HMNS)
is formed, spiral arms are developed in its envelope and continuous
heating occurs due to the collision between the HMNS and spiral arms
(e.g.,~\cite{STU2,KSST,LR}). By this process the maximum temperature
increases to $\sim 30$--50~MeV, and hence, copious neutrinos are
emitted~\cite{Ruffert,Setiawan,Dessart09}.  
To accurately study the evolution of the
hot HMNS with a physical modeling, we have to incorporate
microphysical processes such as neutrino emission and equation of
state (EOS) based on a theory for the high-density and
high-temperature nuclear matter.  However, such simulations have not
been done yet in full general relativity (but see~\cite{OJM} for a
work in an approximate general relativistic gravity with
finite-temperature EOSs).  Incorporation of microphysical processes is
in particular important for exploring the merger hypothesis of SGRB
because it may be driven through pair annihilation of
neutrino-antineutrino pairs~\cite{GRB-BNS}.

In this letter, we present the first results of numerical-relativity
simulation for the BNS merger performed incorporating both a
finite-temperature (Shen's) EOS~\cite{Shen} and neutrino
cooling~\cite{Sekig}. In the following, we report the possible outcome
formed after the merger, criterion for the formation of HMNS and black
hole (BH), thermal properties of the HMNS, and neutrino luminosity
from the HMNS and in the BH formation.

{\em Setting of numerical simulations}: Numerical simulations in full
general relativity are performed using the following formulation and
numerical schemes.  Einstein's evolution equations are solved in the
so-called BSSN-puncture formulation~\cite{BSSN}. We employ the
original version of the geometric variables for the BSSN formulation
together with an improved definition of the conformal factor.  As
in~\cite{KSST}, the dynamical gauge condition for the lapse function
and shift vector is chosen; a fourth-order-accurate finite
differencing in space and a fourth-order Runge-Kutta time integration
are used; a conservative shock capturing scheme with third-order
accuracy in space and fourth-order accuracy in time is employed for
solving hydrodynamic equations. In addition to the ordinary
hydrodynamic equations, we solve evolution equations for the neutrino 
($Y_{\nu}$), electron ($Y_e$), and total lepton ($Y_l$)
fractions per baryon, and, taking into
account weak interaction processes~\cite{Sekig}. Shen's EOS,
tabulated in terms of the rest-mass density ($\rho$), temperature
($T$), and $Y_e$ or $Y_{l}$, is employed.  In addition, we incorporate
neutrino cooling, employing a general relativistic
leakage scheme~\cite{Sekig}. In our leakage scheme,
electron neutrinos ($\nu_e$), electron antineutrinos ($\bar \nu_e$),
and other types ($\mu/\tau$) of neutrinos ($\nu_x$) are taken into
account.

Numerical simulations are performed preparing a non-uniform grid as in
\cite{KSST}. The inner domain is composed of a finer uniform grid and
the outer domain of a coarser nonuniform grid. The grid resolution in
the inner zone is chosen so that the major diameter of each neutron
star in the inspiral orbit is covered by 60 and 80 grid points for
low- and high-resolution runs, respectively: We always performed
simulations for both grid resolutions to confirm that convergence is 
approximately achieved.
Outer boundaries are located in a local wave zone 
(at $\approx 560$--600~km along each coordinate axis which is longer than
gravitational wavelength in the inspiral phase).  During the
simulations, we check the conservation of baryon rest-mass, total
gravitational mass (Arnowitt-Deser-Misner mass plus radiated energy of
GWs), and total angular momentum (including that
radiated by GWs), and find that the errors are within
0.5\%, 1\%, and 3\%, respectively, for the high-resolution runs within
the duration $\approx 30$~ms.

Shen's EOS, derived from a relativistic mean-field theory~\cite{Shen},
is a stiff one among other EOSs, giving the maximum mass of 
zero-temperature spherical neutron stars 
$M_{\rm max} \approx 2.2M_{\odot}$.  
The latest discovery of a high-mass neutron star with mass 
$1.97 \pm 0.04M_{\odot}$~\cite{twosolar} suggests that stiff EOSs are 
favored, and Shen's EOS satisfies this requirement. 
There are two possible fates of BNS~\cite{STU2}: If its total mass 
$M$ is larger than a critical mass $M_c$, a BH will be
formed soon after the onset of the merger, while a differentially
rotating HMNS will be formed for $M < M_c$.  
The value of $M_c$ depends strongly on the EOS.  
Because Shen's EOS is quite stiff, $M_c$ is much larger than the typical 
total mass of BNS, $\sim 2.7M_{\odot}$~\cite{Stairs}, as shown below.  
Thus, with this EOS, a HMNS is the frequent outcome, as in the cases of 
stiff EOSs with which $M_{\rm max}>2M_{\odot}$~\cite{hotoke}.

This paper focuses on the merger of equal-mass BNS with three masses
for each neutron star: $M_{\rm NS}=1.35$, 1.5, and $1.6M_{\odot}$
($M_{\rm NS}$ is the gravitational mass of a neutron star in
isolation). We refer to each model as models L (light), M (middle),
and H (heavy).  We perform the simulation with the initial condition
of about 3--4 orbits before the onset of the merger until the system
relaxes to a quasistationary state. Quasiequilibrium states of BNS are
prepared as the initial conditions, as in~\cite{STU2,KSST}.

\begin{figure}[t]
\epsfxsize=2.9in
\leavevmode
\epsffile{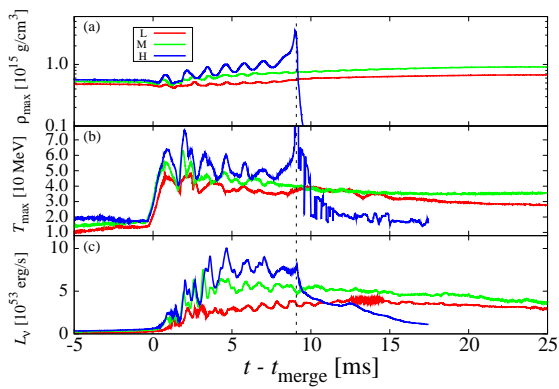}
\vspace{-3mm}
\caption{Maximum rest-mass density, maximum matter temperature, and
total neutrino luminosity as functions of time for three models.
The dashed vertical line shows the time at which a BH is formed 
for model H. 
\label{fig0}}
\end{figure}

\begin{figure*}[t]
\vspace{-2mm}
\epsfxsize=5.4in
\leavevmode
\epsffile{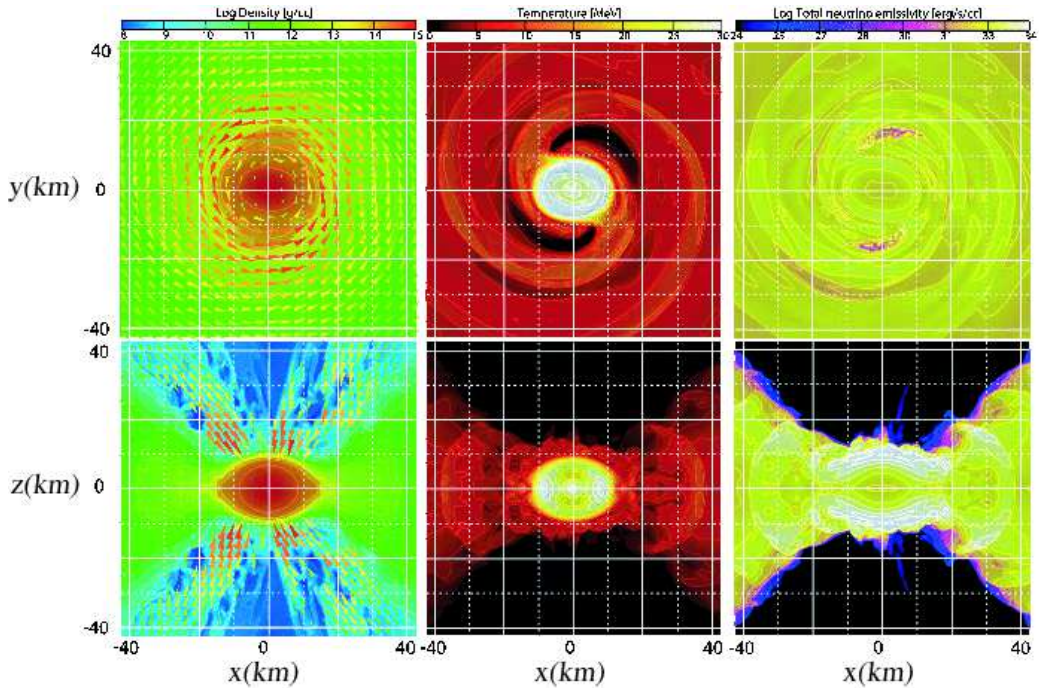}
\vspace{-3mm}
\caption{
Color maps of rest-mass density (with velocity fields), 
temperature, and total neutrino
luminosity at $t \approx 15$~ms after the merger for model M. The
upper and lower panels show the configuration in the $x$-$y$ and
$x$-$z$ planes, respectively.
\label{fig2}}
\end{figure*}

{\em Numerical results}: Figure~\ref{fig0} plots the evolution of the
maximum rest-mass density, $\rho_{\rm max}$, maximum matter
temperature, $T_{\rm max}$, and total neutrino luminosity as functions
of $t-t_{\rm merge}$ where $t_{\rm merge}$ is the onset time of the
merger. For $t< t_{\rm merge}$, $\rho_{\rm max}$ is
approximately constant, while $t\agt t_{\rm merge}$, a HMNS is
formed and subsequently contracts by emission of GWs,
which carry energy and angular momentum from the HMNS; $\rho_{\rm
max}$ increases in the gravitational radiation timescale.  However,
at $t-t_{\rm merge} \sim 20$ ms for models L and M,
the degree of its nonaxial symmetry becomes low enough that the
emissivity of GWs is significantly reduced.  
Because no dissipation process except for neutrino cooling is 
present, the HMNS will be alive at least for the cooling timescale before
collapsing to a BH (see below). 
For model H, the HMNS eventually collapses to a BH after the gradual 
contraction due to the GW emission and a massive disk of $\approx 0.1 M_{\odot}$
is formed around the BH.

The evolution of $T_{\rm max}$ plotted in Fig.~\ref{fig0}(b) shows
that HMNS just after the formation are hot with $T_{\rm max} \sim
50$--70~MeV. Such high temperature is achieved due to the liberation
of kinetic energy of the orbital motion at the collision of two
neutron stars. Subsequently, $T_{\rm max}$ decreases due to
the neutrino cooling, with the maximum luminosity 3--$10
\times 10^{53}$~erg/s (see Fig.~\ref{fig0}(c)), but relaxes to a high
value with $25$--50~MeV. Around the HMNS, spiral arms are formed and
shock heating continuously occurs when the spiral arms hit the HMNS
(cf. Fig.~\ref{fig2}). Due to this process and because of a long
neutrino cooling timescale (see below), the temperature (and thermal
energy) does not significantly decrease in $\sim 10$--100~ms.

Figure~\ref{fig2} plots the color maps of rest-mass density, matter
temperature, and total neutrino luminosity for model M at 
$t-t_{\rm merge}=15$~ms, at which it relaxes to a semi-final
quasisteady state. This shows that the HMNS is weakly spheroidal and
the temperature is high in its outer region. The neutrino luminosity
is also high in its outer region, in particular, near the polar
surface. With the fact that the rest-mass density is relatively small
near the rotation axis above the polar surface, this is a favorable
feature for the merger hypothesis of SGRB; pair annihilation of
neutrinos and antineutrinos could supply a large amount of thermal
energy which may drive a fire ball along the rotation axis.
The pair annihilation efficiency has been approximately estimated in
the previous works~\cite{Ruffert,Setiawan,Dessart09}. These show that
the efficiency for $\nu_{e}\bar{\nu}_{e}$ annihilation is
$\sim 10^{-2}(L_{\nu_{e}}/10^{53}{\rm erg/s})$. If this result holds 
in our work, the pair annihilation luminosity would be 
$\sim 10^{51}$ erg/s.

The reasons that HMNS are formed are; (i) it is rapidly rotating with
the period $\sim 1$~ms, and hence, the centrifugal force increases the
possible mass that can be sustained; (ii) because it is hot, 
thermal energy enhances the pressure. We find that the
rotational velocity with the period $\sim 1$~ms does not play a
substantial role.  Exploring in detail Shen's EOS for high density
tells us that the effect of the thermal energy is significant and can
increase $M_{\rm max}$ by $\sim 20$--30\% for a high-temperature state
with $T \agt 20$~MeV.  Therefore, the HMNS will alive before collapsing to a
BH for a long cooling time $E_{\rm th}/L_{\nu}\sim 2$--3~s where $E_{\rm th}$ 
is total thermal energy of the HMNS.
At the time when the HMNS collapses to a BH, it will be close to a spherical
configuration with low temperature due to longterm GW and neutrino emissions.
Thus, observable signals from the late-time collapse will be weak.

\begin{figure}[t]
\epsfxsize=2.9in
\leavevmode
\epsffile{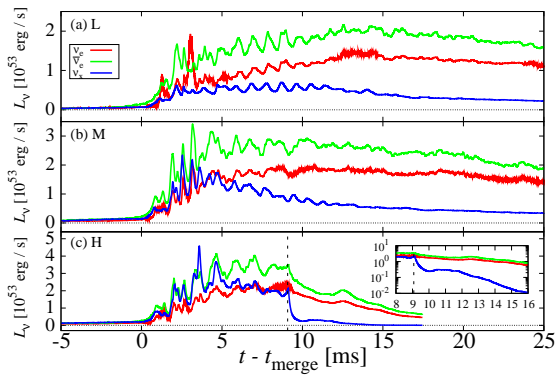}
\vspace{-3mm}
\caption{Neutrino luminosities for three flavors for three models. 
The inset of the bottom panel focuses on the luminosities in 
the BH formation. The meaning of the dashed line is the same as 
in Fig.~\ref{fig0}. 
\label{fig3}}
\end{figure}

Figure~\ref{fig3} plots neutrino luminosities as functions of time for
three flavors ($\nu_{e}$, $\bar{\nu}_{e}$, and sum of $\nu_{x}$). 
It is found that electron antineutrinos are dominantly emitted
for any model. The reason for this is as follows: The HMNS has a high
temperature, and hence, electron-positron pairs are efficiently
produced from thermal photons, in particular in its envelope.  
Neutrons efficiently capture the positrons to emit antineutrinos
whereas electrons are not captured by protons as frequently as
positrons because the proton fraction is much smaller.
Such hierarchy in the neutrino luminosities was reported 
also in \cite{Ruffert}.

Soon after the BH formation for model H, $\mu/\tau$ neutrino luminosity 
steeply decreases because high temperature regions are swallowed into
the BH, while luminosities of electron neutrinos and antineutrinos decrease
only gradually because these neutrinos are emitted via charged-current processes
from the massive accretion disk. We here note that magnetic fields, which are not 
taken into account in the present simulations, could be amplified significantly 
in the accretion disk~\cite{LR} and may play a role in the late evolution of 
the BH-disk system.


The {\em antineutrino} luminosity for the long-lived HMNS is $L_{\bar
\nu} \sim 1.5$--$3\times 10^{53}$~erg/s with small time variability. 
It is by a factor of $\sim 1$--5 larger than that from protoneutron stars 
formed after supernovae~\cite{SN}.  Averaged
neutrino energy is $\epsilon_{\bar\nu} \sim 20$--30~MeV.
The sensitivity of water-Cherenkov neutrino detectors such as
Super-Kamiokande and future Hyper-Kamiokande (HK) have a good
sensitivity for such high-energy neutrinos in particular for electron
antineutrinos~\cite{SM2009}.  The detection number for electron
antineutrinos is approximately estimated by $\sigma \Delta T
L_{\bar\nu}/(4\pi D^2 \epsilon_{\bar\nu})$ where $\sigma$ is the total
cross section of the detector against target neutrinos, $\Delta T$ is
the lifetime of the HMNS, and $D$ is the distance to the HMNS. 
For a one-Mton detector such as HK, the expected detection number 
is $\agt 10$ for $D \alt 5$~Mpc with $\Delta T\sim 2$--3~s, based on an
analysis of~\cite{SM2009}.  Thus, if the BNS merger fortunately happens 
within $D \sim 5$~Mpc, neutrinos from the HMNS may be detected 
and its formation may be confirmed.
Note that GWs from the HMNS will be simultaneously detected for 
such a close event (see below), reinforcing the confirmation of the 
HMNS formation.

\begin{figure*}[t]
\epsfxsize=2.9in
\leavevmode
\epsffile{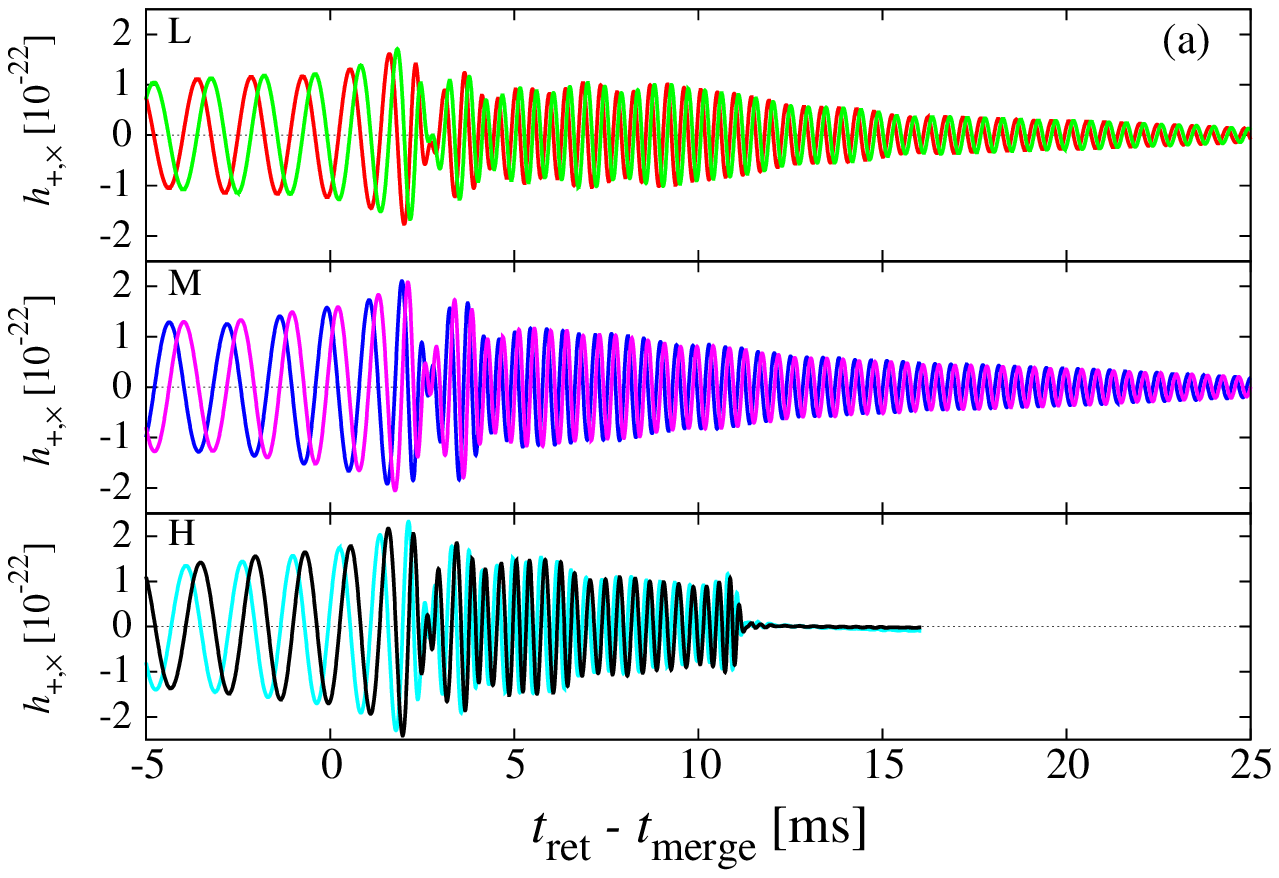}
\epsfxsize=2.9in
\leavevmode
~~~~~~~~\epsffile{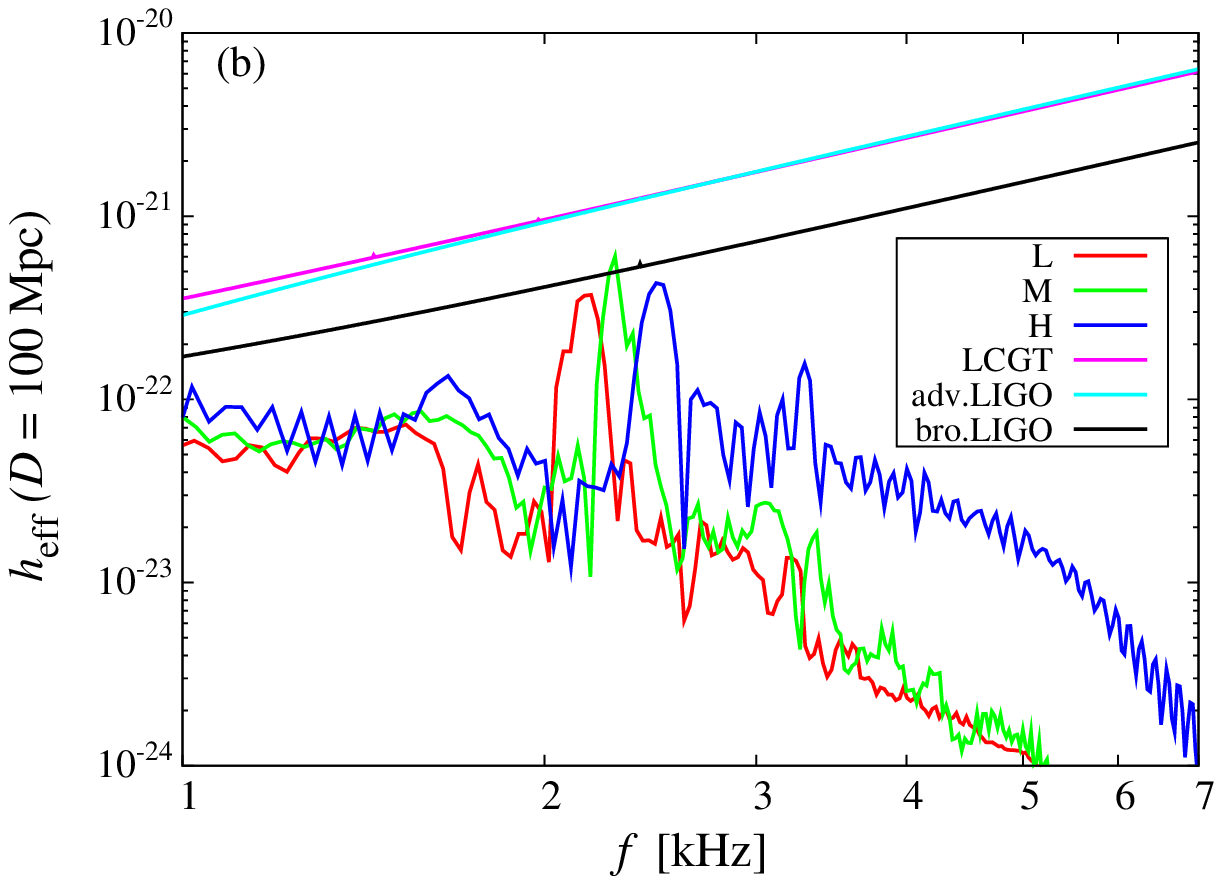}
\vspace{-2mm}
\caption{(a) GWs observed along the 
axis perpendicular to the orbital plane for the hypothetical distance
to the source $D=100$~Mpc. (b) The effective amplitude of
GWs as a function of frequency for $D=100$~Mpc.
The noise amplitudes of Advanced Laser Interferometer 
Gravitational wave Observatories (adv. LIGO), broadband configuration of 
Advanced LIGO (bro. LIGO), 
and Large-scale Cryogenic Gravitational wave Telescope (LCGT)
are shown together.
\label{fig4}}
\end{figure*}

Figure \ref{fig4}(a) plots gravitational waveforms as a function of
retarded time $t_{\rm ret}=t-D-2M {\rm log}(D/M)$ for three models
where $M=2M_{\rm NS}$. Here, $h_+$ and $h_{\times}$ denote the plus
and cross modes of GWs extracted from the metric in
the local wave zone. The waveforms are composed of the so-called chirp
waveform, which is emitted when the BNS is in an inspiral motion (for
$t_{\rm ret} \alt t_{\rm merge}$), and the merger waveform (for
$t_{\rm ret} \agt t_{\rm merge}$), on which we here focus.  For the
HMNS formation, the merger waveforms are composed of quasiperiodic
waves for which $h \alt 10^{-22}$ for $D=100$~Mpc and the peak
frequencies are in a narrow range $f_{\rm peak}=2.1$--2.5~kHz depending weakly on $M$.
They agree with that in the approximate general relativistic 
study~\cite{OJM}. Note that $f_{\rm peak}$ depends on adopted EOS~\cite{OJM}, 
and we will describe the dependence of GWs on EOS elsewhere.
The accumulated effective amplitude, $h_{\rm eff}\equiv
0.4h(f\Delta T)^{1/2}$, is much larger where the factor 0.4 comes from
the averages of angular direction of the source and rotational axis of
the HMNS. Figure \ref{fig4}(b) shows the effective amplitude defined
by $0.4h(f)f \sim 4$--$6 \times 10^{-22}$ for $D=100$~Mpc, where
$h(f)$ is the absolute value of the Fourier transformation of $h_+ + i
h_{\times}$.  This suggests that for a specially-designed version of
advanced GW detectors such as broadband LIGO, 
which has a good sensitivity for a high-frequency band, 
GWs from the HMNS oscillations may be detected with S/N$=5$ if
$D \alt 20$~Mpc or the source is located in an optimistic
direction.

{\em Summary}: We have reported the first results of the
numerical-relativity simulation performed incorporating both a
finite-temperature (Shen's) EOS and neutrino cooling effect.
We showed that for such a stiff EOS, HMNS is the canonical outcome 
and BH is not promptly formed after the onset of the merger as long as the
total mass of the system is smaller than $3.2M_{\odot}$. The primary
reason is that thermal pressure plays an important role for sustaining
the HMNS.  We further showed that the lifetime
of the formed HMNS with mass $\alt 3M_{\odot}$ is much longer than its
dynamical timescale, $\gg 10$~ms, and will be determined by the timescale 
of neutrino cooling.
Neutrino luminosity of the HMNS was shown to be high as $\sim 3$--
$10\times 10^{53}$~erg/s.
The effective amplitude of GWs is 4--$6 \times
10^{-22}$ at $f_{\rm peak}=2.1$--2.5~kHz for a source distance of 100~Mpc. 
If the BNS merger happens at a relatively short source distance or is 
located in an optimistic direction, such GWs may be detected and HMNS 
formation will be confirmed.

{\em Acknowledgments}: Numerical simulations were performed on SR16000
at YITP of Kyoto University and on SX9 and XT4 at CfCA of NAOJ.  This work was
supported by Grant-in-Aid for Scientific Research (21018008, 21105511,
21340051, 22740178), by Grant-in-Aid for Scientific Research on
Innovative Area (20105004), and HPCI Strategic Program of Japanese
MEXT.

\end{document}